\documentclass[journal,twocolumn]{IEEEtran}
\usepackage[cmex10]{amsmath}
\usepackage{amssymb}
\usepackage{cite}
\usepackage{graphicx}
\usepackage{array,color}
\usepackage{multirow}
\usepackage{amsmath}
\usepackage{stfloats}
\usepackage{graphicx}
\usepackage{subfigure}
\usepackage{tabularx}
\usepackage{epsfig,epsf,color,balance,cite}
\usepackage{setspace}
\usepackage{bm}
\usepackage{textcomp}
\usepackage{algorithmic}
\usepackage{algorithm}
\usepackage{comment}
\usepackage{subfigure}
\usepackage{caption}
\usepackage{graphicx}
\usepackage{setspace}
\usepackage{diagbox}
\usepackage{float}

\begin{document}

\title{Channel Estimation for RIS-aided mmWave Massive MIMO System Using Few-bit ADCs }

\author{Ruizhe Wang, Hong Ren, Cunhua Pan, Jun Fang, Mianxiong Dong and Octavia A. Dobre, \IEEEmembership{Fellow, IEEE}
\thanks{R. Wang, H. Ren and C. Pan are with National Mobile Communications Research Laboratory, Southeast University, Nanjing, China. (e-mail:{rzwang, hren, cpan}@seu.edu.cn).
Jun Fang is with the National Key Laboratory
of Science and Technology on Communications, University of Electronic
Science and Technology of China, Chengdu, China (e-mail:junfang@uestc.edu.cn).
Mianxiong Dong is with the Department of Sciences
and Informatics, Muroran Institute of Technology, Muroran, Japan
(e-mail: mx.dong@csse.muroran-it.ac.jp).
Octavia A. Dobre is with the Department of Electrical and Computer Engineering, Memorial University, St. John¡¯s, NL A1B 3X5, Canada (e-mail: odobre@mun.ca).

\emph{Corresponding author: Hong Ren and Cunhua Pan.}}
}

\maketitle


\begin{abstract}
Millimeter wave (mmWave) massive multiple-input multiple-output (massive MIMO) is one of the most promising technologies for the fifth generation and beyond wireless communication system. However, a large number of antennas incur high power consumption and hardware costs, and high-frequency communications place a heavy burden on the analog-to-digital converters (ADCs) at the base station (BS). Furthermore, it is too costly to equipping each antenna with a high-precision ADC in a large antenna array system. It is promising to adopt low-resolution ADCs to address  this problem. In this paper, we investigate the cascaded channel estimation for a mmWave massive MIMO system aided by a reconfigurable intelligent surface (RIS) with the BS equipped with few-bit ADCs. Due to the low-rank property of the cascaded channel, the estimation of the cascaded channel can be formulated as a low-rank matrix completion problem. We introduce a Bayesian optimal estimation framework for estimating the user-RIS-BS cascaded channel to tackle with the information loss caused by quantization. To implement the estimator and achieve the matrix completion, we use efficient bilinear generalized approximate message passing (BiG-AMP) algorithm. Extensive simulation results verify that our proposed method can accurately estimate the cascaded channel for the RIS-aided mmWave massive MIMO system with low-resolution ADCs.


\end{abstract}


\IEEEpeerreviewmaketitle
%
\section{Introduction}
The sixth-generation wireless network is required to have 10-1000 times the capacity of the fifth-generation network, and to be able to serve trillions of devices rather than the current billions of devices \cite{6G}. Millimeter wave (mmWave) massive multiple-input multiple-output (massive MIMO) is a promising technology that exploits the huge available mmWave bandwidth (30-300 GHz) and utilizes the space resources provided by multiple antennas \cite{mmwaveMIMO}. The scale of the antenna array aperture can be significantly decreased due to the short wavelength of mmWave frequencies, making it suitable for small cellular short range communications \cite{mmwaveMIMO2}. Meanwhile, massive MIMO can provide a large antenna array gain, which can help overcome the severe attenuation of mmWave signals and significantly improve the spectrum efficiency.

However, the power consumption and hardware cost considerably rise as a result of the huge bandwidth and large antenna array in mmWave massive MIMO systems. Huge bandwidth requires ultra-high sampling rate analog-to-digital converters (ADCs). Meanwhile, as the sampling rate grows to 100 Msamples per second, the power consumption of ADC increases proportionately \cite{ADC1}. Besides, an excessive number of antennas increases the hardware costs. For large antenna array, the cost of equipping every antenna with a high-precision ADC is unacceptable. Equipping low-resolution ADCs in large antenna array is an efficiency solution, which can significantly lower power consumption and hardware costs. Hence, it is economical and practical to employ low-resolution ADCs for large antenna array and mmWave systems \cite{wenchaokai,mojianhua}.

Furthermore, the mmWave signals suffer from severe path loss and are vulnerable to obstructions. To address this problem, it is promising to deploy reconfigurable intelligent surface (RIS) in mmWave systems \cite{RISmmwave}, where the RIS offers an alternative communication link when the channel between the BS and the user is blocked \cite{RISoverview}. To obtain the performance gain promised by the RIS, the channel state information (CSI) of the cascaded channel should be known accurately. In this paper, we study the cascaded channel estimation for an RIS-aided mmWave massive MIMO system with few-bit ADCs ($\leq 4$ bit). The quantization process causes severe information loss and the traditional channel estimation methods in \cite{zhougui,RISmmwave} based on infinite precision ADCs are no longer applicable for the scenario with low-resolution ADCs. In \cite{junfang}, the low-rank structure of the cascaded channel was explored for channel estimation without considering the quantization effect. In \cite{LS}, the least squares (LS) method was employed to estimate the MIMO channel. However, this method leads to high estimation error since the quantization distortion was simply modeled as additive white Gaussian noise (AWGN). Besides, the authors of \cite{1bitCS} adopted the LASSO algorithm to solve the compressive sensing problem with the few-bit quantizer. However, this method requires long training sequence and the performance is still not good.

In this letter, we adopt the efficient approximate message passing (AMP) algorithms for cascaded channel estimation for mmWave massive MIMO systems aided by an RIS with low-precision ADCs. We formulate the channel estimation problem as a quantized noisy low-rank matrix completion problem. To tackle the information loss caused by the quantization process, we introduce the Bayesian optimal estimator in \cite{wenchaokai} for the estimation of quantizer output. Due to the low-rank property of the cascaded channel matrix, we use modified bilinear generalized approximate message passing (BiG-AMP) algorithm in \cite{BiGAMP} for channel estimation. As performance indicator, we employ the normalized mean square error (NMSE). Simulation results demonstrate that the proposed algorithm outperforms the other traditional estimation algorithms.

{\textit{Notations}}: The symbols $\mathbf{A}^{\ast}$, $\bf{A}^{\text T}$, ${\bf{A}}^{H}$ and $\|{\bf A}\|_{F}$ represent the conjugate, transpose, Hermitian (conjugate transpose) and Frobenius norm of matrix $\bf A$, respectively. The symbols $\mathbb{E}(\cdot)$, $\text {Var}(\cdot)$, $\delta(\cdot)$ and $\mathbb{C}$ represent the expectation operator, the variance operator, the Dirac delta function, and the complex field, respectively. The symbol $a_{ij}$ refers to the $(i,j)$th entry of matrix $\bf A$. The $n\times n$ identity matrix is represented by ${\bf I}_n$. The circularly-symmetric complex Gaussian distribution is denoted as $\mathcal{CN}({\bm{\mu}}, {\bf C})$, where ${\bm{\mu}}$ is the mean vector and ${\bf C}$ is the covariance matrix, respectively.

\section{System Model}\label{systemmodel}

\begin{figure}[H]
\begin{minipage}[t]{0.99\linewidth}
\centering
\includegraphics[width=3.2in]{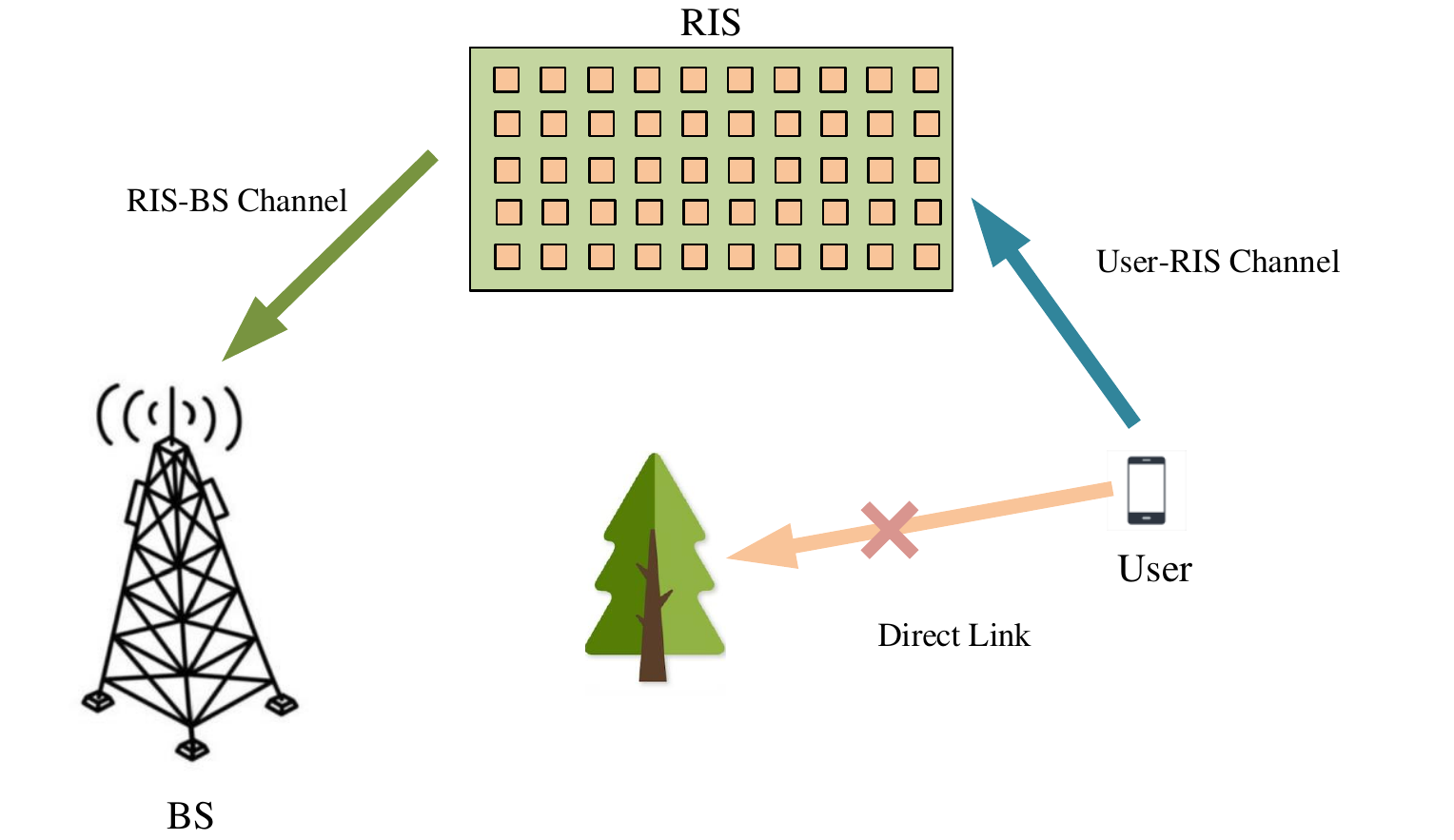}
\vspace{-0.2cm}
\caption{An RIS-aided uplink mmWave massive MIMO system.}
\label{fig0}\vspace{-0.7cm}
\end{minipage}%
\hfill
\end{figure}

We consider a narrow-band mmWave massive MIMO uplink system, as shown in Fig.~\ref{fig0}. An RIS that is composed of $M$ reflecting elements is deployed. The BS is equipped with $N$ transmit antennas to serve a single-antenna user. In this letter, the channels are considered to be quasi-static and block-fading, where the channels in each coherence block stay constant. Denote ${\bf h}\in {\mathbb C}^{M\times 1}$ as the channel from the user to the RIS. The channel between the RIS and the BS is represented by ${\bf G}\in {\mathbb C}^{N \times M}$. The phase shifts of the RIS at time slot $t$ are represented by ${\bf e}_t$, which satisfies $\left|[{\bf e}_t]_m\right|^2=1$ for $1\leq m \leq M$.

At time slot $t$, the signal received at the BS is expressed as
\begin{align}
{\bf y}[t] &={\bf G}{\text {Diag}}({\bf e}_t){\bf h}x_t + {\bf w}_t\nonumber\\
&={\bf G}{\text {Diag}}({\bf h}){\bf e}_tx_t + {\bf w}_t,
\end{align}
where $x_t \in {\mathbb{C}}$ is the transmitted signal, and ${\bf w}_t \in \mathbb{C}^{N\times 1}$ is the additive noise vector at time slot $t$ following the distribution of $\mathcal{CN}\left({\bf 0},\sigma^2_w{\bf I}_{N}\right)$ with $\sigma^2_w$ being the noise power.

By collecting the received signals of $\tau$ time slots, the received signal matrix ${\bf Y}=\left[{\bf y}[1],\cdots,{\bf y}[\tau]\right]$ at the BS is expressed as
\begin{equation}\label{bigY}
{\bf Y}={\bf G}{\text {Diag}}({\bf h}){\mathbf E} + {\bf W} \in {\mathbb{C}^{N\times \tau}},
\end{equation}
where $\mathbf{E}=[{\bf e}_1x_1,\cdots,{\bf e}_{\tau}x_\tau]\in \mathbb{C}^{M\times \tau}$. Each antenna at the receiver side has two ADCs to individually quantize the real and imaginary components of the received signal. After the quantization process, the received signal is written as
\begin{equation}\label{quantizeY}
\widetilde{\bf Y}=\mathcal{Q}\left(\bf Y\right)=\mathcal{Q}\left({\bf U}{\mathbf E}+{\bf W}\right),
\end{equation}
where $\mathcal{Q}(\cdot)$ denotes the quantization function and ${\bf U} = {\bf G}{\text {Diag}}({\bf h}) \in {\mathbb{C}}^{N\times M}$ represents the cascaded channel. The real and imaginary parts of each element of matrix $\bf Y$ are quantized independently.

\subsection{Few-bit Quantization Model}
In this letter, the uniform mid-rise quantization is adopted at the BS, and each complex-valued quantizer has two real-valued $B$-bit quantizers. Denote $\Delta$ as the stepsize of the quantizer. It is assumed that each real-valued quantizer has $2^B-1$ thresholds, denoted as $[r_1,r_2,\cdots,r_{2^B-1}]$, where $r_b=(-2^{B-1}+b)\Delta, 1\leq b \leq 2^B-1$. The quantizer output is set to $r_b-\frac{\Delta}{2}$ when the quantizer input lies between $r_{b-1}$ and $r_{b}$. Without loss of generality, the quantized output of a complex-valued scalar $\xi$ is given by
\begin{align}\label{quantize}
\tilde{y}=&\mathcal{Q}(\xi)\nonumber\\
=&{\text{sign}}\left({\text{Re}(\xi)}\right)\left(\min\left(\left\lceil\frac{|{\text{Re}(\xi)}|}{\Delta_{\text{Re}}}\right\rceil,2^{B-1}\right)-\frac{1}{2}\right)\Delta_{\text{Re}}\nonumber\\
&+{\text j}\;{\text{sign}}\left({\text{Im}(\xi)}\right)\left(\min\left(\left\lceil\frac{|{\text{Im}(\xi)}|}{\Delta_{\text{Im}}}\right\rceil,2^{B-1}\right)-\frac{1}{2}\right)\Delta_{\text{Im}},
\end{align}
where  $\Delta_{\text{Re}}=\left(\mathbb{E}[|{\text{Re}}(\xi)|^2]\right)^{\frac{1}{2}}\Delta$ and $\Delta_{\text{Im}}=\left(\mathbb{E}[|{\text{Im}}(\xi)|^2]\right)^{\frac{1}{2}}\Delta$. The values of the stepsize $\Delta$ are given in \cite{mojianhua}. Specifically, when one-bit quantization is adopted, (\ref{quantize}) takes the following form
\begin{align}
\tilde{y}&={\text{sign}}\left({\text{Re}}(\xi)\right)\sqrt{\frac{2}{\pi}}\left(\mathbb{E}\left[|{\text{Re}(\xi)}|^2\right]\right)^{\frac{1}{2}}\nonumber\\
&\qquad+{\text{sign}}\left({\text{Im}}(\xi)\right)\sqrt{\frac{2}{\pi}}\left(\mathbb{E}\left[|{\text{Im}(\xi)}|^2\right]\right)^{\frac{1}{2}}.
\end{align}
In practice, before the ADCs, a variable gain amplifier (VGA) with an automated gain control (AGC) is utilized, and the measurement of the average power $\mathbb{E}[|{\text{Re}}(\xi)|^2]$ and $\mathbb{E}[|{\text{Im}}(\xi)|^2]$ is employed. It is assumed that $\xi$ has the property of circular symmetry, and thus, $\mathbb{E}[|{\text{Re}}(\xi)|^2]=\mathbb{E}[|{\text{Im}}(\xi)|^2]=\frac{1}{2}\mathbb{E}[|\xi|^2]$ and $\Delta_{\text{Re}}=\Delta_{\text{Im}}$.

\subsection{Cascaded Channel Model}
In this letter, an uniform linear array (ULA) is deployed at both the BS and the RIS. The mmWave channels ${\bf G}$ and ${\bf h}$ are expressed as
\begin{align}
{\bf G}&=\sum_{l=1}^{L}\alpha_l{\bf a}_{N}(\varphi_l){\bf a}^H_{M}(\phi_l),\label{H}\\
{\bf h}&=\sum_{j=1}^{J}\beta_j{\bf a}_{M}(\psi_j),\label{h}
\end{align}
where $L$ and $J$, respectively, represent the number of propagation pathways between the BS and the RIS and the RIS and the user; ${\bf a}_{N}(\varphi_l)\in \mathbb{C}^{N\times 1}$ and ${\bf a}_{M}(\psi_j)\in \mathbb{C}^{M\times 1}$ are the array steering vectors of the ULA at the BS and the RIS, respectively. The coefficients $\alpha_l,\forall l$ and $\beta_j,\forall j$ vary at each channel coherence block. From (\ref{H}) and (\ref{h}), the RIS-aided cascaded channel is modeled as
\begin{equation}\label{cascadedchannel}
{\bf U}= \sum_{l=1}^L\sum_{j=1}^J\alpha_l\beta_j{\bf a}_N(\varphi_l){\bf a}^H_{M}(\phi_l-\psi_j).
\end{equation}

This model illustrates the low-rank property of the cascaded channel matrix for an RIS-aided mmWave communication system. Thus, the cascaded channel estimation can be formulated as a low-rank matrix completion problem.

\section{Channel Estimation Algorithm}
Based on the observation $\widetilde{\bf Y}$ from the few-bit quantizer and the predetermined training matrix $\bf E$, we aim to estimate the cascaded channel $\bf U$ in (\ref{bigY}). Since the quantization process will lose information, we introduce the framework of Bayesian inference to estimate the cascaded channel based on the observation of the quantizer output and use the sum-product algorithm to implement the Bayesian optimal estimators. Due to the low-rank property of the cascaded channel matrix, the cascaded channel estimation can be formulated as a low-rank matrix completion problem. Thus, the modified BiG-AMP algorithm \cite{BiGAMP} is adopted to estimate the cascaded channel ${\bf U}$.


\subsection{Matrix Factorization}
The posterior probability can be calculated as follows according to the Bayes' rule:
\begin{equation}\label{prior}
p({\bf U,E}|\widetilde{{\bf Y}})=\frac{ p(\widetilde{{\bf Y}}|{{\bf U},{\bf E}})p({\bf U})p({\bf E})}{p(\widetilde{\bf Y}{})},
\end{equation}
where
\begin{align}
p({\bf U})&=\prod^N_{n=1}\prod^M_{m=1}p(u_{n,m}|g_{n,m},h_{m})p(g_{n,m})p(h_m)\nonumber\\
&=\prod^N_{n=1}\prod^M_{m=1}\delta(u_{n,m}-g_{n,m}h_{m})p(g_{n,m})p(h_m)
\end{align}
and
\begin{equation}\label{E}
p({\bf E})=\prod^M_{m=1}\prod^\tau_{t=1}\delta(e_{m,t}-\bar{e}_{m,t}).
\end{equation}
In (\ref{E}), $\bar{e}_{m,t}$ is the $m$th row and $t$th column element from a known training matrix $\overline{\bf E}$. To apply the BiG-AMP algorithm, the independent Gaussian priors \cite{hezhengqing} for the RIS-BS channel $\bf G$ and for the user-RIS channel $\bf h$ are assumed, i.e.,
\begin{align}
p({\bf G})&=\prod^N_{n=1}\prod^M_{m=1}p(g_{n,m})=\prod^N_{n=1}\prod^M_{m=1}\mathcal{CN}(g_{n,m};0,\sigma_g^2)\\
p({\bf h})&=\prod^M_{m=1}p(h_m)=\prod^M_{m=1}\mathcal{CN}(h_{m};0,\sigma_h^2),
\end{align}
where $\sigma_g^2$ and $\sigma_h^2$ are the average variances of the RIS-BS channel matrix $\bf G$ and the user-RIS channel vector $\bf h$, respectively. Define ${\bf Z}={\bf U}\overline{\bf E}$, the likelihood $p(\widetilde{\bf Y}|{\bf Z})$ is written as
\begin{align}
p(\widetilde{\bf Y}|{\bf Z})&\triangleq\prod^N_{n=1}\prod^M_{m=1}{p(\tilde{y}_{n,m}|z_{n,m})}\nonumber\\
&=\prod^N_{n=1}\prod^M_{m=1}\left(\frac{1}{\sqrt{\pi \sigma_w^2}}\int_{r_{b-1}}^{r_b}e^{-\frac{\left(y_{n,m}-{\text {Re}}(Z_{n,m})\right)^2}{\sigma_w^2}}{\text d}y\right)\nonumber\\
&\quad\qquad \times \left(\frac{1}{\sqrt{\pi \sigma_w^2}}\int_{r_{b'-1}}^{r_{b'}}e^{-\frac{\left(y_{n,m}-{\text {Im}}(Z_{n,m})\right)^2}{\sigma_w^2}}{\text d}y\right),
\end{align}
when the real output ${\text {Re}}(\widetilde{\bf Y})\in (r_{b-1},r_b]$ and imaginary output ${\text {Im}}(\widetilde{\bf Y})\in (r_{b'-1},r_{b'}]$. The posterior probability for $z$ can be calculated as
\begin{equation}
p(z|\tilde{y})=\frac{p(\tilde{y}|z)p(z)}{p(\tilde{y})},
\end{equation}
where $p(\tilde{y})=\int{p(\tilde{y}|z)p(z){\text d z}}$ is the marginal probability density function (pdf). Suppose $z$ is a complex Gaussian variable with pdf $p(z)=\mathcal{CN}(z;\hat{p},\upsilon^p)$, then, by calculating the integral $\int p(z|\tilde{y})z{\text d}z$, the following posterior mean and variance estimator for $z$ can be obtained
\begin{equation}\label{bayesestimator1}
\hat{z}=\hat{p}+\frac{{\text {sign}}(\tilde{y})\upsilon^p}{\sqrt{2(\sigma^2_w+\upsilon^p)}}\left(\frac{\frac{1}{\sqrt{2\pi}}e^{-\frac{\eta^2_1}{2}}-\frac{1}{\sqrt{2\pi}}e^{-\frac{\eta^2_2}{2}}}{\Phi(\eta_1)-\Phi(\eta_2)}\right)\\
\end{equation}
and
\begin{align}\label{bayesestimator2}
\upsilon^z&=\frac{\upsilon^p}{2}-\frac{{{\left( {{\upsilon }^{p}} \right)}^{2}}}{2\left( \sigma_{w}^{2}+{{\upsilon }^{p}} \right)}\nonumber\\
\times&\left(\frac{\frac{1}{\sqrt{2\pi}}(\eta_1 e^{-\frac{\eta^2_1}{2}}-\eta_2 e^{-\frac{\eta^2_2}{2}})}{\Phi(\eta_1)-\Phi(\eta_2)}+\left(\frac{\frac{1}{\sqrt{2\pi}} (e^{-\frac{\eta^2_1}{2}}- e^{-\frac{\eta^2_2}{2}})}{\Phi(\eta_1)-\Phi(\eta_2)}\right)^2\right),
\end{align}
where $\Phi(x)$ is the cumulative Gaussian distribution function,

\begin{equation}\label{eta1}
\eta_1=\frac{{\text{sign}}(\tilde{y})\hat{p}-\min\{|r_{b-1}|,|r_b|\}}{\sqrt{\frac{\sigma^2_w+\upsilon^p}{2}}},
\end{equation}
and
\begin{equation}\label{eta2}
\eta_2=\frac{{\text{sign}}(\tilde{y})\hat{p}-\max\{|r_{b-1}|,|r_b|\}}{\sqrt{\frac{\sigma^2_w+\upsilon^p}{2}}}.
\end{equation}

Combining (\ref{bayesestimator1}) and (\ref{bayesestimator2}) with (\ref{eta1}) and (\ref{eta2}), we obtain the Bayesian optimal estimator for the real component of $Z$, Due to the circular symmetric property, the estimator of the imaginary component is similar. For the proof of (\ref{bayesestimator1}) and (\ref{bayesestimator2}), please refer to \cite{wenchaokai} for further details.

\subsection{BiG-AMP Algorithm}
The cascaded channel estimation from the output of the quantizer is a quantized and noise-corrupted low-rank matrix reconstruction problem. The proposed BiG-AMP based channel estimation algorithm is presented in Algorithm \ref{alg:4}. In steps 3 and 4, $\hat{p}_{nt}$ and ${\nu^p_{nt}}$ are the estimate of the matrix $\bf Z$ and the corresponding variances, respectively. Steps 3 and 4 differ slightly from the original algorithm in \cite{BiGAMP}, as for a given known training sequence $\overline{\bf E}$, the corresponding variances $\nu^e_{mt}$ are zero. When $\nu^e_{mt}=0$, auxiliary variables that compute the ``plug-in" estimate $\overline{\bf P}=\bar{p}_{nt}$ of the matrix product ${\bf Z}={\bf U}\overline{\bf E}$ and corresponding variances $\bar{\nu}^p_{nt}$ (see \cite{BiGAMP}) were plugged into $\hat{p}_{nt}$ and $\nu^p_{nt}$, respectively. Using the quantities $\widehat{\bf P}$ and ${\nu^p_{nt}}$, the posterior mean $\widehat{\bf Z}$ and variance $\nu^z_{nt}$ are computed in step 5 and 6. The scaled residual $\widehat{\bf S}$ and inverse-residual variances $\nu^s_{nt}$ are calculated in steps 7 and 8 using the posterior moments. Using the residual terms, steps 9 and 10 compute $\widehat{\bf Q}$ and $\nu^q_{mt}$, respectively, where $\hat{q}_{nm}$ can be viewed as an observation  of the cascaded channel matrix $\bf U$ corrupted by $\nu^q_{nm}$-variance-AWGN noise. Finally, steps 11 and 12 estimate the posterior means of $\widehat{\bf U}$ and variance $\nu^u_{nm}$ by using the quantities $\hat{q}_{nm}$ and $\nu^q_{nm}$.
\begin{algorithm}[h]
    \caption{BiG-AMP Based Channel Estimation Algorithm}
    \label{alg:4}
    \begin{algorithmic}[1]
    \REQUIRE $\widetilde{\bf Y}$, $\overline{\bf E}$, $p(\widetilde{\bf Y}|{\bf Z})$, $p({\bf U})$
    \ENSURE $\widehat{\bf U}$
    \STATE Initialize: $i\leftarrow 1; \forall n,t:\hat{s}_{nt}(0)=0, \nu^z_{nt}(0)=1, \hat{z}_{nt}(0)=0; \forall n,m:\nu^u_{nm}(1)=1,\hat{u}_{nm}(1)=0.$
    \FOR{$i =1,\cdots,I_{\text{max}}$}
    \STATE $\forall n,t:{\nu}_{n t}^{p}(i)=\sum_{m=1}^M\nu_{n m}^{u}(i)\left|\bar{e}_{mt}\right|^{2}$
    \STATE $\forall n, t: \hat{p}_{n t}(i)=\sum_{m=1}^M\hat{u}_{nm}(i)\bar{e}_{mt}-\hat{s}_{nt}(i-1) {\nu}_{nt}^{p}(i) $\
    \STATE $\forall n, t: \nu_{n t}^{z}(i)=\operatorname{Var}\left\{z_{n t} \mid p_{n t}=\hat{p}_{n t}(i) ; \nu_{n t}^{p}(i)\right\} $
    \STATE $\forall n, t: \hat{z}_{n t}(i)=\mathbb{E}\left\{z_{n t} \mid p_{n t}=\hat{p}_{n t}(i) ; \nu_{n t}^{p}(i)\right\} $
    \STATE $\forall n, t: \nu_{n t}^{s}(i)=\left(1-\nu_{n t}^{z}(i) / \nu_{n t}^{p}(i)\right) / \nu_{n t}^{p}(i) $
    \STATE $\forall n, t: \hat{s}_{n t}(i)=\left(\hat{z}_{n t}(i)-\hat{p}_{n t}(i)\right) / \nu_{n t}^{p}(i)$
    \STATE $ \forall n, m: \nu_{n m}^{q}(i)=\left(\sum_{t=1}^{\tau}\left|\bar{e}_{m t}(i)\right|^{2} \nu_{n t}^{s}(i)\right)^{-1} $
    \STATE \mbox{$\forall n, m: \hat{q}_{n m}(i)= \hat{u}_{n m}(i)+\nu_{n m}^{q}(i) \sum_{t=1}^{\tau} \bar{e}_{m t}^{*}(i) \hat{s}_{n t}(i)$}
    \STATE \mbox{$\forall n, m: \nu_{n m}^{u}(i+1)= \operatorname{Var}\left\{{u}_{n m} \mid {q}_{n m}=\hat{q}_{n m}(i) ; \nu_{n m}^{q}(i)\right\} $}
    \STATE \mbox{$\forall n,m: \hat{u}_{n m}(i+1)=\mathbb{E}\left\{{u}_{n m}\mid {q}_{n m}=\hat{q}_{n m}(i) ; \nu_{n m}^{q}(i)\right\}$}
    \ENDFOR
    \end{algorithmic}
\end{algorithm}


\subsection{Benchmark Algorithms}
In this section, we describe two benchmark algorithms for signal reconstruction. The first is the LS estimation adopted in \cite{LS}. The LS estimator is given by
\begin{equation}\label{LS}
\widehat{{\bf U}}_{\text{LS}}=\widetilde{\bf Y}\overline{{\bf E}}^{H}\left(\overline{{\bf E}} \overline{{\bf E}}^{H}\right)^{-1},
\end{equation}
where the known matrix $\overline{\bf E}$ has full row-rank.

Another approach is linear minimum mean square (LMMSE) estimation adopted in \cite{mojianhua}. Using the Bussgang's theorem \cite{bussgang}, the quantizer output $\widetilde{\bf Y}$ is decomposed into the signal component and the quantization noise ${\bf W}_q$, which is independent from the signal component as
\begin{align}
{\bf Y}&=\mathcal{Q}\left({\bf U}\overline{{\bf E}}+{\bf W}\right)\nonumber\\
&=(1-\eta_b)\left({\bf U}\overline{{\bf E}}+{\bf W}\right)+{\bf W}_{\text q},
\end{align}
where $\eta_b\triangleq {\mathbb{E}[|\mathcal{Q}(\xi)-\xi|^2]}/{\mathbb{E}[|\xi|^2]}$ is the quantization normalized MSE (NMSE) given in \cite{quantizing}. Define $\widehat{\bf W}=(1-\eta_b){\bf W}+{\bf W}_{\text q}$ as the effective noise matrix. Using the approximate LMMSE (ALMMSE) method in \cite{bussgang}, we obtain the following ALMMSE estimator
\begin{align}\label{ALMMSE}
&\widehat{{\bf U}}_{\text{ALMMSE}}\nonumber\\
&={\widetilde{\bf Y}}{{\mathbf{\overline{E}}}^{H}}{{\left( \left( 1-{{\eta }_{b}} \right)\mathbf{\overline{E}} {{{\mathbf{\overline{E}}}}^{H}}+\left( \left( 1-{{\eta }_{b}} \right)\frac{\sigma _{\omega }^{2}}{\sigma _{u}^{2}}{{\mathbf{I}}_{M}}+{{\eta }_{b}}N{{\mathbf{I}}_{M}} \right) \right)}^{-1}},
\end{align}
where $\sigma_u^2=\frac{\sigma_g^2\sigma_h^2}{\sigma_g^2+\sigma_h^2}$. The expression (\ref{ALMMSE}) is similar to the LS estimator in (\ref{LS}) except that the former has a regularized inverse.

\subsection{Complexity Analysis}
The complexity introduced by basic matrix multiplications in steps 3, 9 and 10 are $\mathcal{O}(NM\tau)$, the computations needed in step 4 to step 8 of the proposed BiG-AMP algorithm are $\mathcal{O}(N\tau)$, and that of the reminding steps of the algorithms is $\mathcal{O}(NM+M\tau)$ \cite{BiGAMP}. Therefore, the overall computational complexity of the proposed BiG-AMP algorithm is at most $I_{\text{max}}\mathcal{O}(NM\tau)$.

\section{Simulation Results}\label{simulation}
In this section, we present simulation results to demonstrate the effectiveness of the proposed algorithm for the cascaded channel estimation of the RIS-aided MIMO systems with few-bit ADCs. The antenna spacing of ULA is half-wavelength for the BS and the RIS. The training matrix $\overline{\bf E}$ is generated from shifted-Zadoff-Chu sequences \cite{mojianhua}. The path gain coefficients $\{\alpha_l\}$ and $\{\beta_j\}$ are generated from $\mathcal{CN}(0,1)$. The angular parameters $\varphi_l$, $\phi_l$ and $\psi_j$ independently follow the uniform distribution of $(0,1]$. The signal-to-noise ratio (SNR) is defined as ${\text{SNR}}\triangleq\frac{\mathbb{E}[\|{\bf Z}\|^2_{\text F}]}{\mathbb{E}[\|{\bf W}\|^2_{\text F}]}=\frac{\mathbb{E}[\|{\bf U}\overline{\bf{E}}\|^2_{\text F}]}{\mathbb{E}[\|{\bf W}\|^2_{\text F}]}$. The number of propagation paths is set as $L=J=10$. The number of antennas employed at the BS is $64$ and the number of reflecting elements at the RIS is $32$. Note that the number of reflecting elements is less than the number of BS antennas and a small $L$ is used due to the fact that the BiG-AMP algorithm requires that the matrix ${\bf U}$ is a tall $(N>M)$ and low-rank matrix. The normalized NMSE is defined as ${\text{NMSE}\left({\bf U}\right)}\triangleq\mathbb{E}\left[\frac{\left\|\widehat{{\bf U}}-{\bf U}\right\|^2_{ F}}{\left\|{{\bf U}}\right\|^2_{ F}}\right]$.

\begin{figure}[ht]
\begin{minipage}[t]{1\linewidth}
\centering
\includegraphics[width=3.2in]{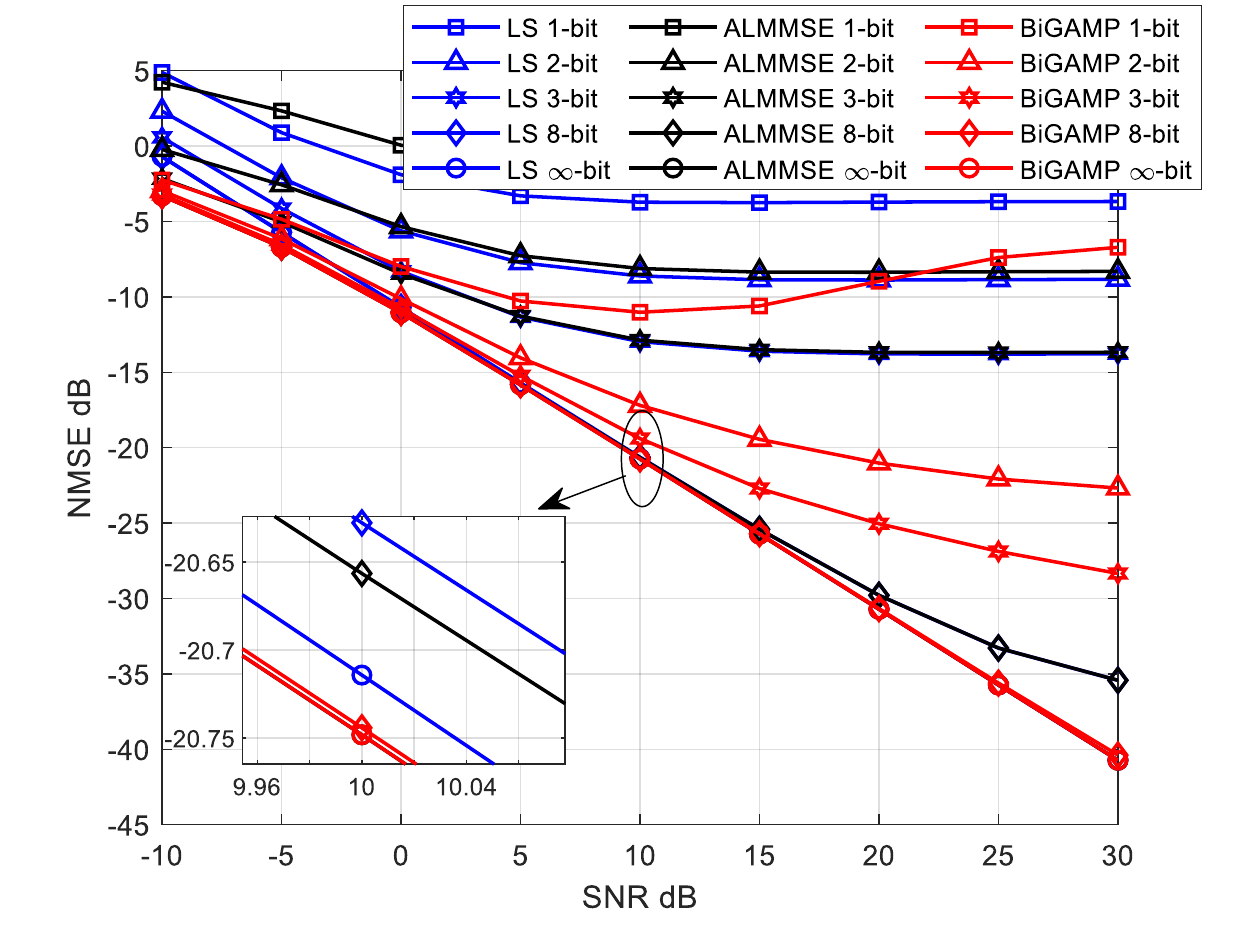}
\vspace{-0.2cm}
\caption{NMSE of {\bf U} versus the SNR, the training length $\tau=$500.}
\label{fig1}\vspace{0cm}
\end{minipage}%
\hfill
\end{figure}

In Fig.~\ref{fig1}, we illustrate the NMSE versus the SNR. The ADC resolutions are set to 1, 2, 3, 8 and infinite bits, respectively. The figure shows that the BiG-AMP algorithm performs significantly better than the LS and ALMMSE algorithms with 1,2 and 3-bits of resolution. As a result, the proposed BiG-AMP algorithm's effectiveness for channel estimation with few-bit ADCs is demonstrated. Besides, it can be observed that the performance of 8-bit quantization
is almost as good as that of infinite-bit quantization. Hence, 8-bits quantization can achieve the performance close to the case
of infinite-bit quantization. In addition, the NMSE decreases with SNR except for the one-bit quantization. The primary cause is that the one-bit quantization involves highly nonlinear processing. At low SNR, the noise may be helpful to improve the distinction for this system. This phenomenon for the one-bit quantization system is known as stochastic resonance \cite{1bit}.
\begin{figure}[ht]
\begin{minipage}[t]{1\linewidth}
\centering
\includegraphics[width=3.2in]{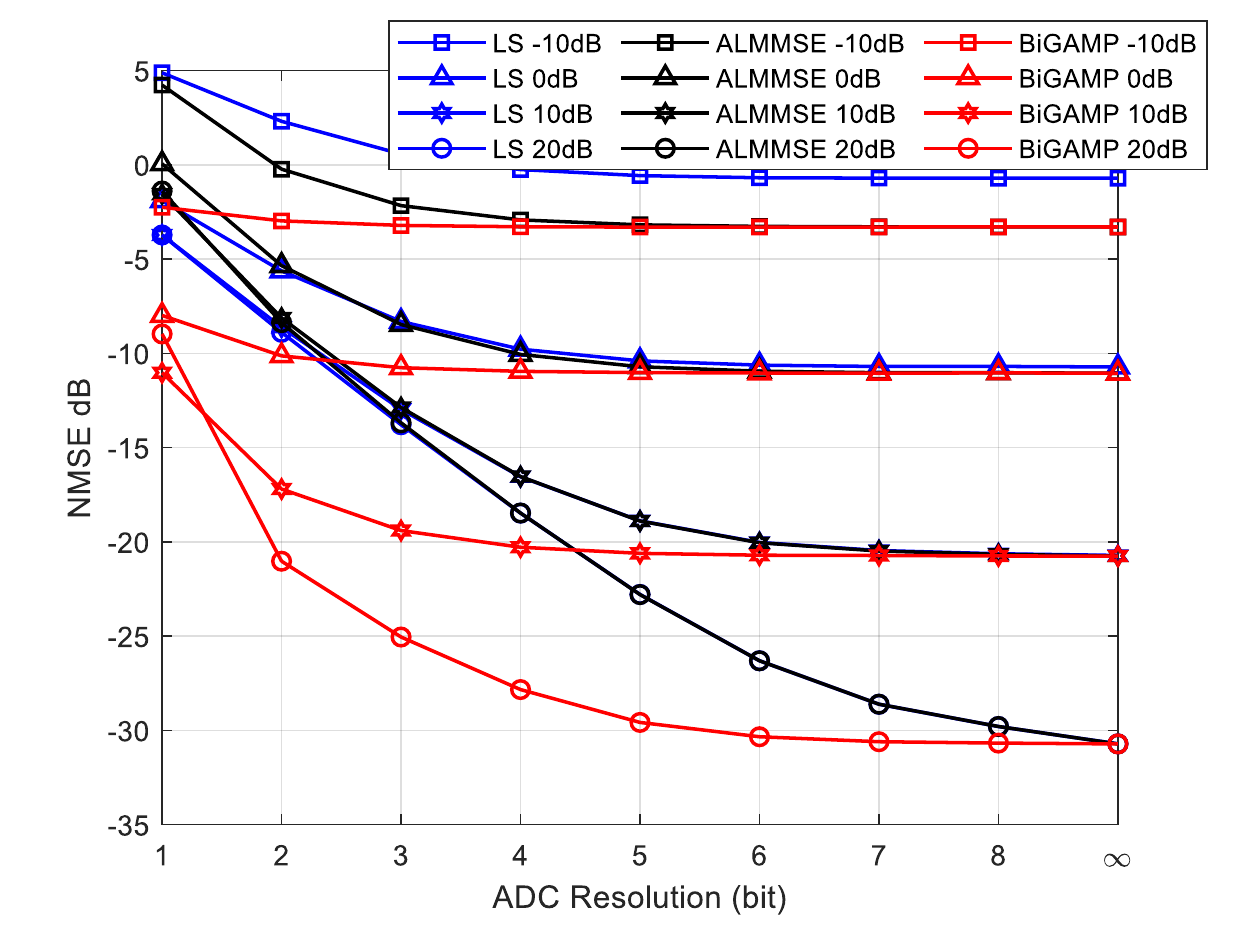}
\vspace{-0.2cm}
\caption{NMSE of {\bf U} versus the ADC resolution, the training length $\tau=$500.}
\label{fig2}\vspace{0cm}
\end{minipage}%
\hfill
\end{figure}

Fig.~\ref{fig2} depicts the NMSE versus the ADC resolutions with SNR $=$ -10, 0, 10 and 20 dB, respectively. From Fig.~\ref{fig2}, it can be observed that the BiG-AMP algorithm performs much better than the LS and ALMMSE algorithms when the ADC resolution is 1, 2 and 3 bits. As the ADC resolution increases, the performance of the three algorithms is similar.

\begin{figure}[ht]
\begin{minipage}[t]{1\linewidth}
\centering
\includegraphics[width=3.2in]{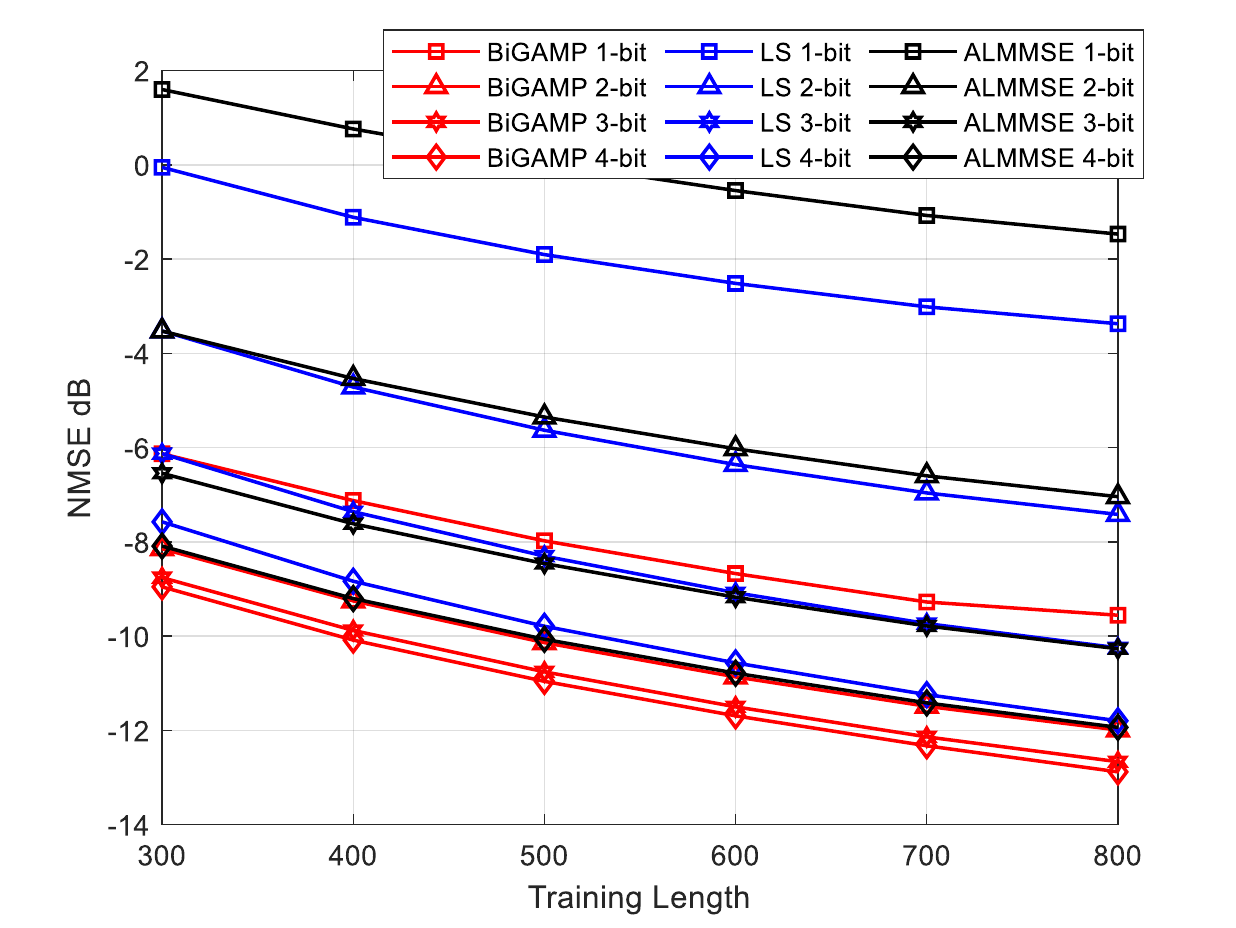}
\vspace{-0.2cm}
\caption{NMSE of {\bf U} versus the training length, the SNR$=0$ dB.}
\label{fig3}\vspace{0cm}
\end{minipage}%
\hfill
\vspace{-0.3cm}
\end{figure}
The NMSE versus the training length $\tau$ is shown in Fig.~\ref{fig3}. Due to the more available measurements when $\tau$ is large, the NMSE decreases with training length for all algorithms.

\section{Conclusion}\label{conclusion}
In this letter, we studied the channel estimation for an RIS-aided mmWave MIMO system with low-precision ADCs. Since the cascaded channel matrix has low-rank property, we formulated the low-rank matrix completion problem for channel estimation. Since the low-resolution quantization causes much information loss, we introduced the Bayesian optimal estimator and proposed to adopt the modified BiG-AMP algorithm to solve the bilinear matrix completion problems that estimate the cascaded channel with known prior information about its distribution. Simulation results demonstrated that the proposed BiG-AMP algorithm outperforms the traditional estimation algorithms.

\bibliographystyle{IEEEtran}
\bibliography{myre}


\end{document}